\documentclass[twocolumn,showpacs,preprintnumbers,amsmath,amssymb]{revtex4}

\usepackage{graphicx}
\usepackage{dcolumn}
\usepackage{bm}

\begin{document}

\preprint{APS/123-QED}

\title{Theory of magnetoresistance based on variable-range hopping in the presence of Hubbard interaction and spin-dynamics}

\author{M. Wohlgenannt} \email{markus-wohlgenannt@uiowa.edu}

\address{Department of Physics and Astronomy and Optical Science and Technology Center, University of Iowa, Iowa City, IA 52242-1479, USA}

\date{\today}

\begin{abstract}

We develop a theory of magnetoresistance based on variable-range hopping. An exponentially large, low-field and necessarily positive magnetoresistance effect is predicted in the presence of Hubbard interaction and spin-dynamics under certain conditions. The theory was developed with the recently discovered organic magnetoresistance in mind. To account for the experimental observation that the organic magnetoresistance effect can also be negative, we tentatively amend the theory with a mechanism of bipolaron formation.

\end{abstract}


\maketitle

\section{\label{sec:Introduction}Introduction}

\subsection{Organic magnetoresistance}

Organic magnetoresistance (OMAR) is a recently discovered large, low field, room-temperature magnetoresistive effect ($\Delta R(B)/R \leq 10\%$ at $B=10 mT$) in organic light-emitting diodes (OLEDs) \cite{Francis:2004,Mermer:2005,Prigodin:2006}. To the best of our knowledge, the mechanism causing OMAR is currently not firmly established, although Prigodin et al. \cite{Prigodin:2006} have put forward a model of OMAR based on spin-dependent carrier recombination and spin-dynamics caused by hyperfine interaction. For the benefit of the reader, we will briefly summarize some of the main experimental results:

\begin{enumerate}
  \item It is an effect associated with the bulk resistance of the organic semiconductor layer \cite{Francis:2004,Mermer:2005} \label{item:bulk}
  \item It is only weakly dependent on the minority carrier density \cite{Sheng:2006}, and occurs also in heavily p-doped devices \cite{Nguyen:2006} \label{item:unipolar}
  \item It is independent of the magnetic field direction \label{item:isotropic}
  \item It obeys the empirical laws $\Delta R(B)/R \propto B^2/(B^2+B_0^2)$ or $\Delta R(B)/R \propto B^2/(|B|+B_0)^2$ dependent on material, where $B_0 \approx 5mT$ in most materials \cite{Mermer:2005PRB} \label{item:law}
  \item The $B_0$ value is much larger in materials with strong spin-orbit coupling \cite{Sheng:2006submitted} \label{item:spinorbit}
  \item It is only weakly temperature dependent \label{item:temperature}
  \item It typically decreases with increasing voltage and carrier density \cite{Francis:2004,Mermer:2005} \label{item:weakdependence}
  \item Its sign can be positive or negative, dependent on material and/or operating conditions of the devices \cite{Mermer:2005PRB} \label{item:sign}
\end{enumerate}

Observation (\ref{item:unipolar}) establishes that OMAR is not related to carrier recombination in distinction to the model put forward by Prigodin. (\ref{item:spinorbit}) shows that OMAR is caused by spin-dynamics.

\subsection{Correlation effects on hopping conductivity}

Although a solid understanding of carrier transport in disordered organic films has not yet been achieved, it is commonly believed that it occurs through variable-range hopping \cite{Kaiser:2001,Vissenberg:1998}. Furthermore, it is known that Coulomb-correlation effects are much more important in organics than in inorganics because of the smaller dielectric constant as well as confinement of the wavefunction to individual molecules. Electron-electron interaction effects on hopping conduction have been studied for many years. For example, the intersite Coulomb interaction \cite{Pollak:1979,Efros:1975} results in the concept of "Coulomb gap". Yamaguchi, Aoki and Kamimura studied the effect of \emph{intrastate} Coulomb interaction (Hubbard-like interaction) in the strongly localized regime \cite{Yamaguchi:1979}. Their model is based on the following Hamiltonian:

\begin{equation}\label{equ:Hamiltonian}
    H=\sum_{\alpha \sigma} \epsilon_{\alpha \sigma} n_{\alpha \sigma} + \frac{1}{2}\sum_{\alpha \sigma} U_{\alpha} n_{\alpha \sigma}n_{\alpha -\sigma}
\end{equation}

\begin{figure}
  \includegraphics[width=\columnwidth]{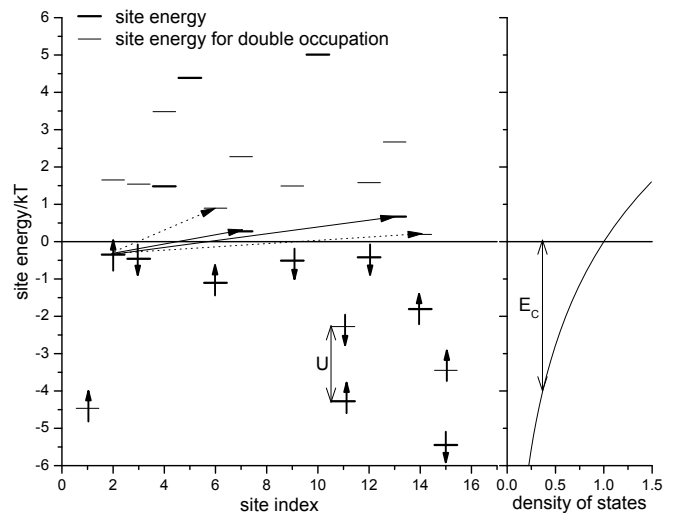}\\
  \caption{Schematic picture of the impurity states (left panel) and their density of states (right panel). Hops are allowed (solid arrows) from SO to all UO sites within the thermal energy kT from the Fermi energy, $E_F=0$, hops between SO sites are allowed only for antiparallel spins. However, in the case of rapid spin-flips hops between other SO sites become allowed (dashed arrows). $E_C$ is an energy characterizing the density of states distribution below $E_F$.}\label{fig:Fig1}
\end{figure}

$\epsilon_{\alpha \sigma}$, $U_{\alpha}$ and $n_{\alpha \sigma}$ represent the one-electron energy, the Hubbard interaction energy and a number operator of state $\alpha$ with spin $\sigma$, respectively. Their model assumes a constant value for $U$ and a uniform distribution for $\epsilon_{\alpha \sigma}$ with a density of states $\nu$. The conduction mechanism is assumed to be a one-electron hop assisted by one phonon. In the presence of the intrastate interaction there exist three electronic states for each Anderson localized state, namely the unoccupied (UO) state, the singly occupied (SO) state and the doubly occupied (DO) state. SO states are found just below the Fermi level, $E_F$, down to $E_F-U$ and carry free spins, whereas DO states involved in hopping, which are spin-singlets, have an occupied deep level located near ($E_F-U$) and accommodate an additional electron of opposite spin in a state located close to $E_F$ (see Fig.~\ref{fig:Fig1}). Accordingly, there exist the following four different kinds of hopping processes: (1) hopping from SO to UO (2) hopping from SO to SO (3) hopping from DO to UO and (4) hopping from DO to SO, where the occupancy refers to that prior to the hop. In our treatment in section~\ref{sec:OMAR} we will neglect hops out of DO sites because of their smaller density of states (see Fig.~\ref{fig:Fig1}). YAK showed that Mott's law for the temperature dependent resistivity, $\rho (T)$ still holds even in the presence of intrastate correlation \cite{Yamaguchi:1979}:

\begin{eqnarray}
  \rho (T) &=& \rho_{\infty} exp \left (\frac{T_0}{T} \right )^\frac{1}{4} \label{equ:rhoMott} \\
  T_0 &=& \frac{13.6}{k\nu \xi^3} \label{equ:rhoMott2}
\end{eqnarray}

where $\xi$ is the localization length and k is Boltzmann's constant, although the value of $\xi$ will differ whether U is considered or not. We have written Mott's law for the three-dimensional case, which may not apply to organic semiconductors. The replacement $1/4\rightarrow 1/(d+1)$ in the exponent must be made in case of a different dimensionality, $d$. Eq.~\ref{equ:rhoMott} is valid only in the low field regime. In OLEDs, however, the applied electric field is quite large, typically $F \approx 10^5 V/cm$. For large fields eq.~\ref{equ:rhoMott} remains valid, if the replacement $T\rightarrow \xi e F/k$ is made, where e is the elementary charge \cite{Apsley:1975}.

\subsection{Magnetoresistance}

Although the intrastate interaction does not change the form of $\rho (T)$, Kamimura and Kurobe showed that $U$ plays an essential role in magnetoresistance \cite{Kamimura:1983}. When a magnetic field is applied, the spins of SO states become parallel to the field. Therefore, the probability of finding a pair of antiparallel spins becomes smaller with increasing magnetic field, and the hopping processes from SO to SO states become suppressed. Such a strong positive magnetoresistance, which saturates at high fields, has indeed been observed experimentally \cite{Jiang:1994,Essaleh:1995}. It has also been measured in strongly localized magnetic two-dimensional electron gases and interpreted using a similar model \cite{Smorchkova:1998}. However, the characteristic magnetic field scale above which the effect saturates is very large, orders of magnitude larger than the 5mT characteristic magnetic field scale of OMAR. Therefore, in this form the Kamimura-Kurobe model cannot account for OMAR. However, this model does not take into account spin-flips. In organic semiconductors, it is well-known that there exist spin-flip mechanisms, the best-known of which is hyperfine interaction \cite{Steiner:1989}. Hyperfine interaction results in the precession of the electron spin around the nuclear spins of the hydrogen atoms, and the precession period is typically 1-10ns. Furthermore, since the drift mobility, $\mu$ in disordered organic semiconductors is typically very low, about $10^{-4} cm^{2} (Vs)^{-1}$, the distance covered by the carrier within the precession period of the spin is only 0.1 to 1nm \cite{Sheng:2006} (see note~\cite{diffusivehopping}), similar to the molecular dimensions. Therefore, it should be necessary to take the spin-precession into account. In the limit of rapid spin-precession, hops become allowed to all SO sites for all carriers, rather than only for ones in a singlet state. Furthermore, since the hyperfine interaction is very weak, upon application of a magnetic field of typically several milliTesla the spin-dynamics is suppressed \cite{Sheng:2006}.

The problem of variable-range hopping for large U, further taking account of the possibility of spin-flips, was considered by Meir and Altshuler \cite{Meir:1997}. The effect of spin-orbit coupling on hopping magnetoconductivity was also investigated by Shapir and Ovadyahu, where it was concluded that backscattering known from weak-localization theory is still important in the strongly localized regime \cite{Shapir:1989}. Meir and Altshuler's calculation was only performed for a large applied magnetic field, which polarizes all SO states, thus blocking all SO to SO hopping processes. This leads to an effectively reduced density of states, and an exponentially enhanced magnetoresistance. Therefore the only possibility of such hops is through a spin-flip with probability $P_{SF}$. Their main result is that the variable-range hopping resistance is given by the parallel circuit of two resistors, $\rho_1$ and $\rho_2$ given through:

\begin{eqnarray}
  \left (ln \frac{\rho_1}{\rho_\infty} \right )^4 &=& \frac{T_0}{T} \label{equ:Meir1} \\
  \left (ln \frac{\rho_2}{\rho_\infty} \right )^4 + \left (ln \left ( \frac{\rho_2}{\rho_\infty} P_{SF} \right ) \right )^4 &=& \frac{2 T_0}{T} \label{equ:Meir2}
\end{eqnarray}

\section{Theory of organic magnetoresistance \label{sec:OMAR}}

Sheng et al. \cite{Sheng:2006} have shown that in organics
\begin{equation}\label{equ:PSF}
    P_{SF}(B)=P_{SF}(0)\frac{B_0^2}{B^2+B_0^2},
\end{equation}
where $B_0$ is the hyperfine coupling strength, or more generally the strength of the spin-flip mechanism. Therefore application of B reduces $P_{SF}$, because the associated Zeeman energy pins the spins, and it is straightforward to show that the theory by Meir and Altshuler can only result in positive magnetoresistance. This is contrary to experimental observations for OMAR, which can be positive or negative. However, we will pursue this theory further. At the end of our treatment, we will suggest a modification to the theory such that it can result in negative magnetoresistance also.

The basic idea underlying the theory of variable-range hopping is that the average hopping distance is directly related to the density of states of potential target sites. Therefore, we need to determine the density of states available to a hop out of a particular site (site 2 in Fig.~\ref{fig:Fig1}). The density of UO target sites is determined by the density of states around the Fermi energy, $\nu (E_F)$, while that of SO target sites is determined by $\nu (E_F-U)$. Without spin-flips the overall density of target sites is $\nu = \nu (E_F)+1/2 \nu (E_F -U)$ since half the SO states will have a spin opposite to that of the hopping carrier. With spin-flips however we have $\nu = \nu (E_F)+\nu (E_F-U )$. Quite generally we can state that $\nu=\nu(U,P_{SF})$. Therefore we assume that Mott's law still holds, but the replacement $\nu \rightarrow \nu(U,P_{SF})$ must be made in eq.~\ref{equ:rhoMott2}. Finally, we note that a similar mechanism should also hold for nearest neighbor hopping, because of the possibility that the nearest neighbor site is an SO site with parallel spin. Then hopping into this site would be forbidden unless rapid spin-flips occur.

\section{Discussion}

To simplify further discussion, we linearize the modified Mott's law to obtain for the magnetoresistance:

\begin{equation}\label{equ:MR}
    \frac{\Delta R(B)}{R}=-\left (\frac{T_0}{T} \right )^\frac{1}{d+1} \frac{1}{d+1}\frac{\Delta \nu (B)}{\nu},
\end{equation}

Furthermore,

\begin{eqnarray}
  \frac{\Delta \nu}{\nu} &=& -\frac{1}{2} \frac{\nu (E_F-U )}{\nu (E_F)+\nu (E_F-U )} \\
  &\approx & -\frac{1}{2} exp \left (-\frac{U}{E_C} \right ),
\end{eqnarray}

where, in the second equation we have made the assumption, as it is commonly done in disordered organics, that $\nu(E)=exp((E-E_F)/E_C)$ \cite{Blom:2000} (Fig.~\ref{fig:Fig1}, right panel), although a Gaussian density of states is commonly believed to be more accurate. If we moreover employ a crude estimate for $T_0/T \approx (E_C/kT)$, then we obtain the following approximate result, which is useful for quantitative estimates and for examining qualitative trends:

\begin{equation}\label{equ:MRestimate}
    \frac{\Delta R(B)}{R} \approx \alpha^\frac{1}{d+1} \frac{1}{2(d+1)} e^{-\frac{U}{E_C}} P_{SF}(0) \frac{B^2}{B^2+B_0^2},
\end{equation}

where $\alpha = E_C/kT$. We would like to remind the reader that $\alpha $ is also the exponent of the current-voltage (I-V) characteristics predicted by the theory of space-charge limited current in the presence of traps, i.e. $I\propto V^{\alpha+1}$ which is routinely applied to OLEDs \cite{Blom:2000}.

Unfortunately, we are not aware of any experimental measurement of $U$. It is important not to confuse $U$ with $U_1 \approx 5 eV$ used routinely in quantum chemical calculations for $\pi$-conjugated molecules \cite{Soos:1984}, which refers to the interaction between electrons located on the same \emph{intramolecular} site. However, we can use $U_1$ to estimate $U \approx U_1 (a/\xi)^2$, where $a \approx 1-5 \AA$ is the size of an intramolecular site. We therefore expect that $U$ is about two orders of magnitude smaller than $U_1$. Finally, we estimate $P_{SF}(0)$, which will be determined by the competition between spin-flips and hops, therefore

\begin{equation}
    P_{SF}(0) \approx \frac{g \mu_B B_0/\hbar}{\mu F/\xi+g \mu_B B_0/\hbar}. \label{equ:PSO0}
\end{equation}

\subsection{Predictions of the model}

\begin{itemize}
\item Since the model is based on hopping conduction, OMAR should not exist in organic single-crystal devices with band transport. To the best of our knowledge, this has not yet been tested
\item The magnitude of OMAR depends critically on the ratio $U/E_C$.
\item Our model is obviously consistent with experimental observations (\ref{item:bulk}), (\ref{item:unipolar}), (\ref{item:isotropic}), (\ref{item:law}), (\ref{item:spinorbit})
\item Our model is an example of a spin-dependent process which does not require thermal spin-polarization, which is in agreement with (\ref{item:temperature})
\item In the high-field case the effect is predicted to decrease with increasing $F$, since $F$ replaces $T$ in eq.~\ref{equ:MR}. Furthermore, $F$ reduces the effect through eq.~\ref{equ:PSO0}. These statements are in agreement with (\ref{item:weakdependence})
\item The effect tends to be smaller in high-mobility materials \cite{Mermer:2005PRB}. Theoretically this is a consequence of eq.~\ref{equ:PSO0}, and because there should also exist an inverse relationship between $\mu$ and $E_C$.
\item Making a guess for $U \approx 0.1 eV$, taking typical values for $E_C \approx 0.1 eV$ \cite{Schmechel:2004} and $\alpha \approx 10$ \cite{Mermer:2005}, assuming rapid spin-flips $P_{SF}=1$ and $d=3$, we obtain as an estimate $\Delta R/R \approx 10\%$, certainly consistent with OMAR
\end{itemize}

\subsection{Speculative modification to achieve negative magnetoresistance}

The main shortcoming of the model is that it predicts only positive magnetoresistance, in contradiction with experimental observation~(\ref{item:sign}). Therefore, a modification of the theory is required. Although we have been unable to find a mechanism that is firmly established, we are able to suggest a candidate mechanism for which some experimental evidence exists. As we described above, spin-flips will lead to an increase in the population of DO (spin 0) sites at the expense of SO (spin 1/2) sites. We propose that bipolaron formation \cite{Heeger:1988} can take place at DO sites. The formation of bipolarons is well-established for heavily doped organic semiconductors \cite{Harima:1999,Appel:1999,Deussen:1992}, evidence in OLEDs is much sketchier \cite{Greenham:1996,Wohlgenannt:2000}. Bipolarons are commonly expected to have lower mobility than polarons, which are well-known to be the spin-1/2 charge carriers in organic semiconductors. This expectation results from the observation that, to account for the binding energy of the bipolarons, the lattice distortion around a bipolaron is greater than that around a polaron, resulting in additional mass and reduced hopping probability. Since the applied magnetic field reduces spin-flips and therefore occupation of DO sites, bipolaron formation will be suppressed, resulting in negative magnetoresistance. More generally, any mechanism through which the correlation length of DO sites becomes less than that of SO sites can result in negative magnetoresistance.

\section{Conclusion}

We have proposed a theory for the recently discovered organic magnetoresistance effect. Our theory is based on the variable range hopping model which states that the average hopping distance is determined by the density of states of potential target sites. We show that, in the presence of Hubbard interaction, this density is restricted by Pauli's principle unless rapid spin-flips occur. In organics, spin-flips are known to result from the hyperfine interaction, which can be eliminated by the application of a weak external field. Therefore the application of an external field leads to an exponential increase in resistance. It has been brought to our attention that Bobbert and coworkers are currently developing a similar model.

\section{Acknowledgements}

We acknowledge many fruitful discussions with M. E. Flatt\'{e}. This work was supported by NSF Grant No. ECS 04-23911.

\bibliography{Bibliography}

\begin{thebibliography}{30}
\bibitem[{dif()}]{diffusivehopping}
\bibinfo{note}{The situation is actually more complicated: The spin-precession
  must be large compared to the overall (diffusive) hopping rate, not just the
  rate of downhill hops which contribute to the drift mobility. However,
  considering the very large electric fields applied in OLEDs, the difference
  should not be too great.}
\end{thebibliography}

\end{document}